\documentclass[11pt]{article}
\usepackage{hyperref}
\pdfoutput=1
\begin{document}
\title{Vortex formation and instability in the left ventricle}
\author{Trung Bao Le and Fotis Sotiropoulos
\\\vspace{6pt} Department of Civil Engineering,\\
University of Minnesota, Minneapolis, MN 55455, USA\\
\\Dane Coffey and Daniel Keefe
\\\vspace{6pt} Department of Computer Science and Engineering\\
 University of Minnesota, Minneapolis, MN 55455, USA}
\maketitle
\begin{abstract}
We study the formation of the mitral vortex ring during early diastolic filling in a patient-specific left ventricle (LV) using direct numerical simulation. The geometry of the left ventricle is reconstructed from Magnetic Resonance Imaging (MRI) data of a healthy human subject. The left ventricular kinematics is modeled via a cell-based activation methodology, which is inspired by cardiac electro-physiology and yields physiologic LV wall motion. In the fluid dynamics videos, we describe in detail the three-dimensional structure of the mitral vortex ring, which is formed during early diastolic filling. The ring starts to deform as it propagates toward the apex of the heart and becomes inclined. The trailing secondary vortex tubes are formed as the result of interaction between the vortex ring and the LV wall. These vortex tubes wrap around the circumference and begin to interact with and destabilize the mitral vortex ring.  At the end of diastole, the vortex ring impinges on the LV wall and the large-scale intraventricular flow rotates in clockwise direction. We show for the first time that the mitral vortex ring evolution is dominated by a number of vortex-vortex and vortex-wall interactions, including lateral straining and deformation of vortex ring, the interaction of two vortex tubes with unequal strengths, helicity polarization of vortex tubes and twisting instabilities of the vortex cores. 
\end{abstract}

\end{document}